\documentclass[aps,preprint, groupaddress, amsmath, amssymb, longbibliography]{revtex4-1}
\usepackage{float} 
\usepackage{graphicx}
\usepackage{dcolumn}
\usepackage{bm}
\usepackage{upgreek}
\usepackage{times}
\usepackage{mathrsfs}
\usepackage{multirow}
\usepackage{fancyhdr} 
\usepackage{hyperref} 
\usepackage{fancyheadings}

\begin{document}

\title{Alloying strategy for two-dimensional GaN optical emitters}

\author{C.~Pashartis}
\author{O.~Rubel}
\email[]{rubelo@mcmaster.ca}
\affiliation{Department of Materials Science and Engineering, McMaster University, 1280 Main Street West,
Hamilton, Ontario L8S 4L8, Canada}

\date{\today}

\begin{abstract}
The recent progress in formation of two-dimensional (2D) GaN by a migration-enhanced encapsulated technique opens up new possibilities for group III-V 2D semiconductors with a band gap within the visible energy spectrum. Using first-principles calculations we explored alloying of 2D-GaN to achieve an optically active material with a tuneable band gap. The effect of isoelectronic III-V substitutional elements on the band gaps, band offsets, and spatial electron localization is studied. In addition to optoelectronic properties, the formability of alloys is evaluated using impurity formation energies. A dilute highly-mismatched solid solution 2D-GaN$_{1-x}$P$_x$ features an efficient band gap reduction in combination with a moderate energy penalty associated with incorporation of phosphorous in 2D-GaN, which is substantially lower than in the case of the bulk GaN. The group-V alloying elements also introduce significant disorder and localization at the valence band edge that facilitates direct band gap optical transitions thus implying the feasibility of using III-V alloys of 2D-GaN in light-emitting devices.
\end{abstract}


\maketitle

\pagestyle{fancy} 
\fancyhead{} 
\rhead{Published in: Phys. Rev. B \textbf{96}, 155209 (2017); DOI: \href{https://doi.org/10.1103/PhysRevB.96.155209}{10.1103/PhysRevB.96.155209}}

%
%

\section{Introduction}

Two-dimensional (2D) materials have demonstrated utility in various technologies such as transistors, photodetectors, and supercapacitors \cite{Mas-Balleste_N_3_2011, Fiori_NN_9_2014, Koppens_NN_9_2014} enabling new functionalities (flexible, transparent electronics) and further device miniaturization \cite{Wang_NN_7_2012,Butler_ACSN_7_2013}. The epitaxial growth of layered materials on heterogeneous surfaces results in van der Waals heterostructures that can accommodate a larger lattice mismatch than traditional bulk heterostructures \cite{Koma_TSF_216_1992,Geim_N_499_2013}. The current portfolio of materials for optoelectronic device applications is dominated by bulk III-V semiconductors. 2D III-V materials feature a wide range of band gaps that span the entire visible spectrum \cite{Blase_PRB_51_1995,Sahin_PRB_80_2009,Zhuang_PRB_87_2013,Balushi_NM_15_2016}, which makes them potentially useful in optoelectronics. Drawbacks of 2D III-V monolayers are their much heavier effective masses, when compared  to their bulk counterparts, an indirect band gap character, and formability issues \cite{Zhuang_PRB_87_2013}.  The experimental realization of 2D III-V's beyond hexagonal-BN is extremely challenging due to the high energy difference of $0.3-0.6$~eV/atom between the 2D and bulk structures \cite{Zhuang_PRB_87_2013}.

\citet{Balushi_NM_15_2016} recently reported the growth of 2D-GaN via a migration-enhanced encapsulated technique with the use of graphene to enclose a bilayer GaN and maintain its structural stability. This pioneering study opens a possibility to overcome a formability issue by manufacturing a few-layer 2D III-V semiconductors. Although the planar monolayer GaN has an indirect band gap of $E_\text{g}\approx4$~eV, a few-layered 2D-GaN is a direct band gap material with a tuneable band gap of $E_\text{g}\approx4-5.3$~eV depending on the layer thickness  \cite{Zhuang_PRB_87_2013, Balushi_NM_15_2016}. The larger band gap of 2D-GaN \textit{vs} the bulk GaN is attributed to quantum confinement effects. The quantum confinement effects are also responsible for an enhanced exciton binding energy $E_\text{ex}$. A binding energy of the order of 1~eV is expected for 2D-GaN due to a linear scaling $E_\text{ex}\sim E_\text{g}/4$ between the band gap and the exciton binding energy of 2D semiconductors \cite{Jiang_PRL_118_2017}. This result for the exciton binding energy subtracted from the values of the band gap energy implies that blue-ultraviolet emissions are possible with 2D-GaN, whereas transition metal dicalcholgenides exhibit lower optical transition energies of $1\!-\!2$~eV \cite{Ramasubramaniam_PRB_84_2011,Wang_NN_7_2012,Butler_ACSN_7_2013} even though the true band gaps are substantially higher because of the strong excitonic effects \cite{Hanbicki_SSC_203_2015}. The blue-ultraviolet energy spectrum of 2D-GaN raises a question about possibility of alloying the material with other III-V elements to tune the emission wavelength across the visible spectrum. In this Letter, we probe the effects of isoelectronic substitutions on spatial electron localization, changes in band gap energies, band edge offsets as well as solubility of impurities in the host 2D-GaN utilizing the density functional theory (DFT) \cite{Kohn_PR_140_1965}.


\section{Method}

Vienna \textit{ab-initio} simulation program (VASP) \cite{Kresse_PRB_54_1996, Kresse_CMS_6_1996} DFT package was employed in this work. PBEsol \cite{Perdew_PRL_100_2008} was used as the exchange-correlation functional since it accurately captures structural properties and the strength of GaN chemical bonds. The calculated formation enthalpy of $\Delta H_\text{f}=-1.29$~eV  for wurtzite GaN is in a good agreement with the range of experimental values $-1.63$~eV \cite{Ranade_JPCB_104_2000} and $-1.34$~eV \cite{Jacob_JCG_311_2009}. Projector-augmented wave  potentials \cite{Blochl_PRB_50_1994} were used with  $spd$ and $sp$ valence electrons for group-III and V elements, respectively. The Brillouin zone of the 2D-GaN primitive unit cell was sampled with a $8\times8\times1$ k-mesh. The full structural optimization was performed with an energy convergence criteria of $10^{-6}$~eV and the plane-wave cutoff energy of 500~eV. The resultant lattice constant of the 2-atom primitive planar GaN cell corresponded to 3.18~{\AA} after relaxation. Each 2D structure had a fixed vacuum spacing of 20~$\text{\r{A}}$ to ensure the monolayer nature of the system.

Impurities were modelled using a supercell technique. We substituted a single group-III/V element into its associative isoelectronic site in a host GaN 128-atom planar $8\times8$ supercell corresponding to $\approx1.6$\% dilution. Supercell calculations were undertaken with the corresponding downscaled k-mesh. Internal relaxation of atomic positions due to impurities was performed using a force convergence of 0.01~eV/$\text{\r{A}}$  while maintaining the macroscopic lattice parameters of the host 2D-GaN. Relativistic effects (the spin orbit interaction) were simulated for supercells with In, Tl, Sb, and Bi impurities. The K and $\Gamma$-points were examined as they correspond to valence and conduction band edges.  Presence of a large vacuum space enables band offset calculations to within 30~meV of an error for heavier elements via a direct comparison of the band edge eigenvalues, i.e., without additional techniques to account for a reference energy shift. Born-effective charge calculations were done using both the self-consistent response to a finite electric field and perturbative \cite{Gajdos_PRB_73_2006} approaches due to numerical issues. Calculations for the bulk wurtzite GaN were performed using $8\times8\times4$ k-mesh for the primitive Brillouin zone. Impurities were introduced in a 128-atom supercell of the size $4\times4\times2$ with a Brillouin zone sampled using a $2\times2\times2$ k-mesh. Crystallographic information  files (CIF) with atomic structures used in calculations can be accessed through the Cambridge crystallographic data centre (CCDC deposition numbers 1579497-1579514, 1579732).

\section{Results and Discussion}

\begin{figure}
	\includegraphics[width=0.9\columnwidth]{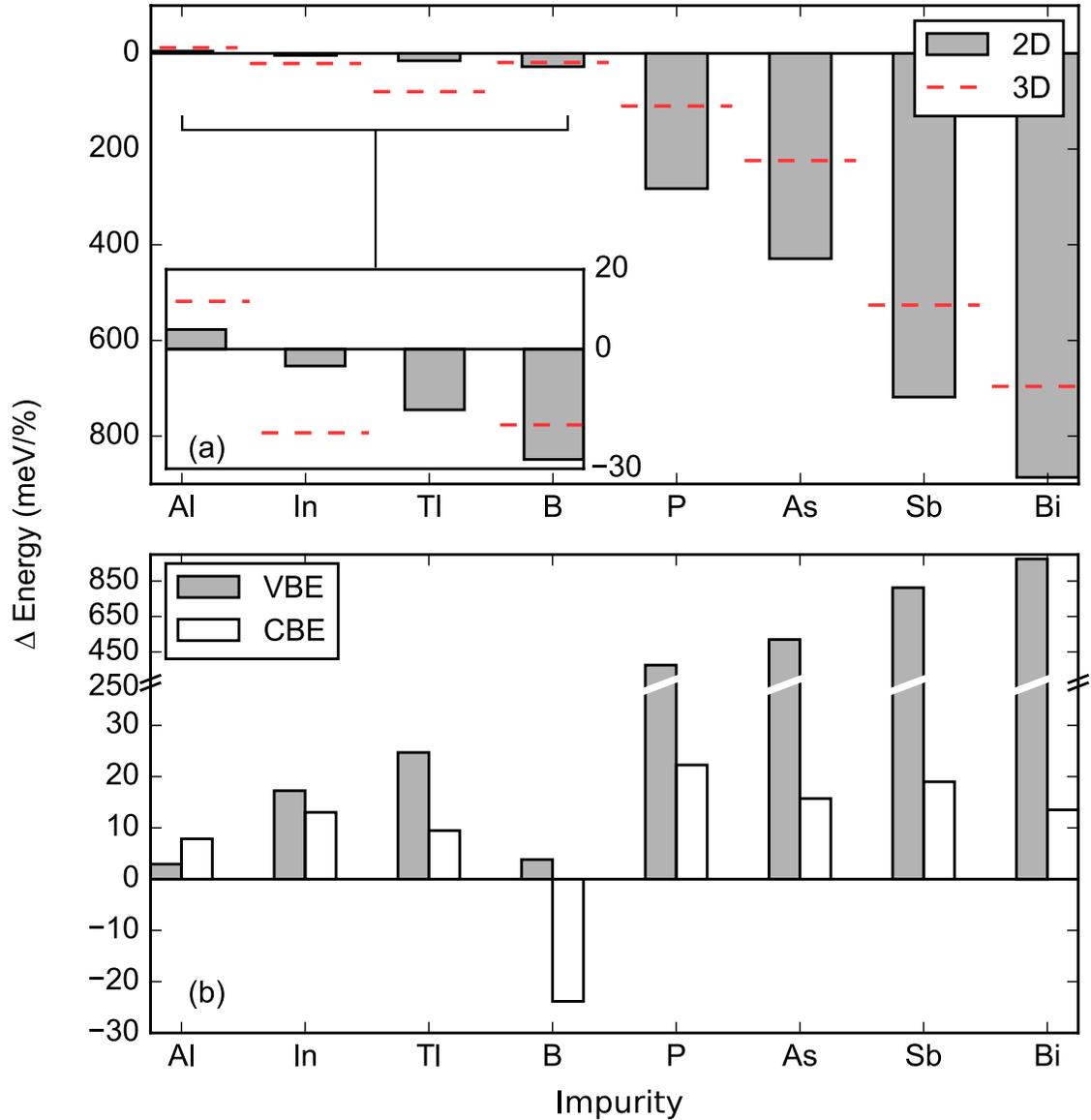}
	\caption{ (a) The band gap change per percent concentration of III-V substitutional isoelectronic elements in 2D-GaN and bulk GaN. (b) Respective shifts of the valence and conduction band edges (VBE and CBE) relative to the host 2D-GaN.}
	\label{FIG:bandgaps}
\end{figure}

We begin by presenting the results of the band gap changes due to incorporation of individual isoelectronic impurities into the host 2D-GaN structure as shown in Fig.~\ref{FIG:bandgaps}a. The group-III elements do not change the band gap significantly with boron reducing the band gap the most in a dilute limit. In contrast, group-V elements affect the band gap much more significantly. Figure~\ref{FIG:bandgaps}a includes data for the bulk wurtzite GaN. Notably changes of the band gap due to impurities in the bulk correlate with that in the monolayer. 

Results for the bulk GaN can be compared to experimental values of the band gap variation taken linearly up to $x=10$\% concentration at room temperature: 1~meV/\% for Al$_x$Ga$_{1-x}$N \cite{Brunner_JAPL_82_1997,Lee_APL_74_1999}  and $-25$~meV/\% for In$_x$Ga$_{1-x}$N \cite{Shan_APL_69_1996, Matsuoka_APL_7_2002}, which are consistent with our results. Furthermore, previous theoretical calculations of wurtzite GaN$_{1-x}$P$_x$ structures yield a band gap variation of $-150$~meV/\% to 10\% of P \cite{Bi_APL_69_1996}, $-170$~meV/\% for 6\% of As in  GaN$_{1-x}$As$_x$ \cite{Tan_AIPA_5_2015}, and $-400$~meV/\% for 10\% of Sb in GaN$_{1-x}$Sb$_x$ \cite{Sheetz_PRB_84_2011} that show similar trends to our calculations. The theoretical calculations capture experimental trends, however quantitatively the calculated band gap variation may contain errors ranging from $5-20$~meV/\% for group-III elements to $50-100$~meV/\% for group-V elements. Possible reasons for discrepancy include alloy statistics and clustering effects that are not captured in the calculation of supercells with a single impurity atom.

The band offsets due to impurities in 2D-GaN are shown in Fig.~\ref{FIG:bandgaps}b. Remarkably all impurities raise the valence band edge (VBE) more than the conduction band edge (CBE) with exception of boron and aluminum. The group-V elements have a much stronger effect on the VBE when substituted in 2D-GaN as compared to group-III elements. Variations in the Born effective charge of impurities shown in Table~\ref{TABLE:born_eff_loc} aid in explaining this effect. In general, perturbations in the valence band scale with the electropositivity of the substituted element (increase in $Z^* - Z^*_\text{host}$). The shift of the valence band edge due to group-V impurities can be viewed as introducing a trap state in the vicinity of the VBE. Similarly, substituting nitrogen into a 2D-GaAs host lattice produces perturbations to the CBE, which correlates with N being more electronegative element than As. This feature is not unique to 2D-GaN and was previously reported for the bulk GaN and GaAs \cite{Mattila_PRB_58_1998,Bellaiche_PRB_54_1996,OReilly_SST_24_2009,Pashartis_PRA_7_2017}. Dilute borides seem to provide the only possibility for engineering a conduction band offset as it is the most electronegative element that can be introduced in the host. Note that the estimated inaccuracy in the band offsets is less than 5~meV/\% for group-III elements and 20~meV/\% for group-V elements, which is due to the subtle changes in the average potential of the supercell introduced by impurities.

\begin{table}
	\caption{Born effective charges of impurity elements in the host 2D-GaN lattice and localization characteristics at the band edges evaluated using the IPR ($\chi$). The effective charge is highly anisotropic and listed for the parallel ($\parallel$) and perpendicular ($\perp$) directions with respect to the 2D-GaN plane.}
\label{TABLE:born_eff_loc}
	\begin{ruledtabular}
		\begin{tabular}{l c c c c}
			Impurity & $Z^*_\parallel$ & $Z^*_\perp$ & \multicolumn{2}{c}{$\chi$}\\
			\cline{4-5}
			element & & & VBE & CBE\\
		\hline
		B & 2.67 & 0.45 & 0.0170 & 0.0079 \\
		Al & 2.82 & 0.45 & 0.0129 & 0.0079 \\
		In & 2.80 & 0.45 & 0.0127 & 0.0078  \\
		Tl & 2.82 & 0.45 & 0.0137 & 0.0078 \\
		P & $-$1.93 & $-$0.29 & 0.2614 & 0.0079 \\
		As & $-$1.74 & $-$0.28 & 0.4016 & 0.0079\\
		Sb & $-$1.38 & $-$0.28 & 0.2849 & 0.0080 \\
		Bi & $-$1.18 & $-$0.24 & 0.4015 & 0.0080 \\
		ref. 2D-GaN & $\pm$ 3.04& $\pm$ 0.33 & 0.0127 & 0.0078 
		\end{tabular}
	\end{ruledtabular}
\end{table}

The lack of a direct band gap in the planar 2D-GaN requires changes in the band dispersion in order to facilitate optical transitions. Localized states due to the impurities  relax the momentum conservation requirement and thus enable direct optical transitions. An inverse participation ratio (IPR) criterion that quantifies localization across the structure was introduced by  \citet{Wegner_ZPBCM_36_1980}. It can be viewed as a statistical method that determines a normalized variance of the wavefunction constructed from second moments. Here we use a discrete version of IPR \cite{Murphy_PRB_83_2011} that is evaluated on the basis of probabilities $\rho_\alpha$ of finding an electron with an eigenenergy $E_i$ within the muffin tin spheres centred at atomic sites $\alpha$
\begin{equation}
   \chi (E_i) = \dfrac
   {\sum_{\alpha}^N \rho_\alpha^2 (E_i)}
   {\left[\sum_{\alpha}^N \rho_\alpha (E_i) \right]^2}~.
\end{equation}
The summation index $\alpha$ runs over all atomic sites. Here the participation ratio $\chi^{-1}$ represents a number of atomic sites, which confine the wave function $\psi_i(\bm{r})$ with $\chi=1$ being the upper limit of IPR. The lower limit of IPR, $\chi=1/N$, is the inverse number of atoms in the lattice that corresponds to extended Bloch states.

\begin{figure}
	\includegraphics[width=1\columnwidth]{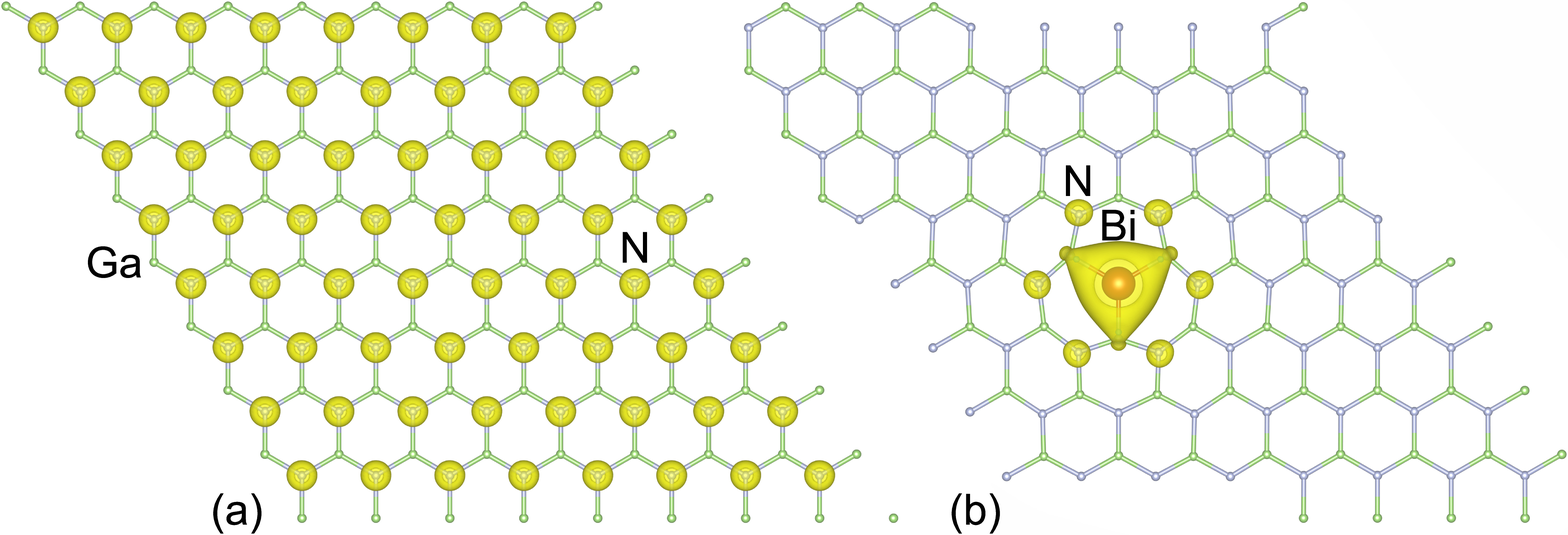}
	\caption{Isosurfaces of the wave function probability amplitude $|\psi_{n,\mathbf{k}}(\mathbf{r})|^2$ at the valence band edges for (a) Ga$_{64}$N$_{64}$, (b) Ga$_{64}$N$_{63}$Bi$_1$ using VESTA \cite{Momma_JAC_44_2011}. Both plots share the same isovalue of 0.00038~Bohr$^{-3}$ for both spins. A large shift in the isosurface is observed to center around Bi.}
	\label{FIG:charge_density_isosurfaces}
\end{figure}

The IPRs in our structures at the band edges can be found in Table~\ref{TABLE:born_eff_loc}. The CBE is insignificantly affected by any of the substitutional elements, even when their incorporation is accompanied by a significant lattice strain due to the size mismatch as in the case of Bi and Tl. The VBE demonstrates a minor increase in localization for group-III elements since Ga and the substitutional elements are electronically similar in the presence of N, as evident from the similar values of the Born effective charges (Table~\ref{TABLE:born_eff_loc}). The group-V elements introduce strongly localized states in the valence band (an order of magnitude larger IPR), which can be attributed to their much more electropositive nature than N by more than a single elementary charge in $Z^*_\parallel$. The localization can be visually observed in Fig.~\ref{FIG:charge_density_isosurfaces}. The VBE isosurfaces are centered at the N-sites in the host structure but reside almost entirely on the substituted Bi-atom in GaN:Bi. Chemical trends in localization characteristics of III-V impurities in 2D-GaN are very similar to those previously reported by \citet{Bellaiche_PRB_54_1996} for the bulk GaN.

\begin{figure}
	\includegraphics[width=0.9\columnwidth]{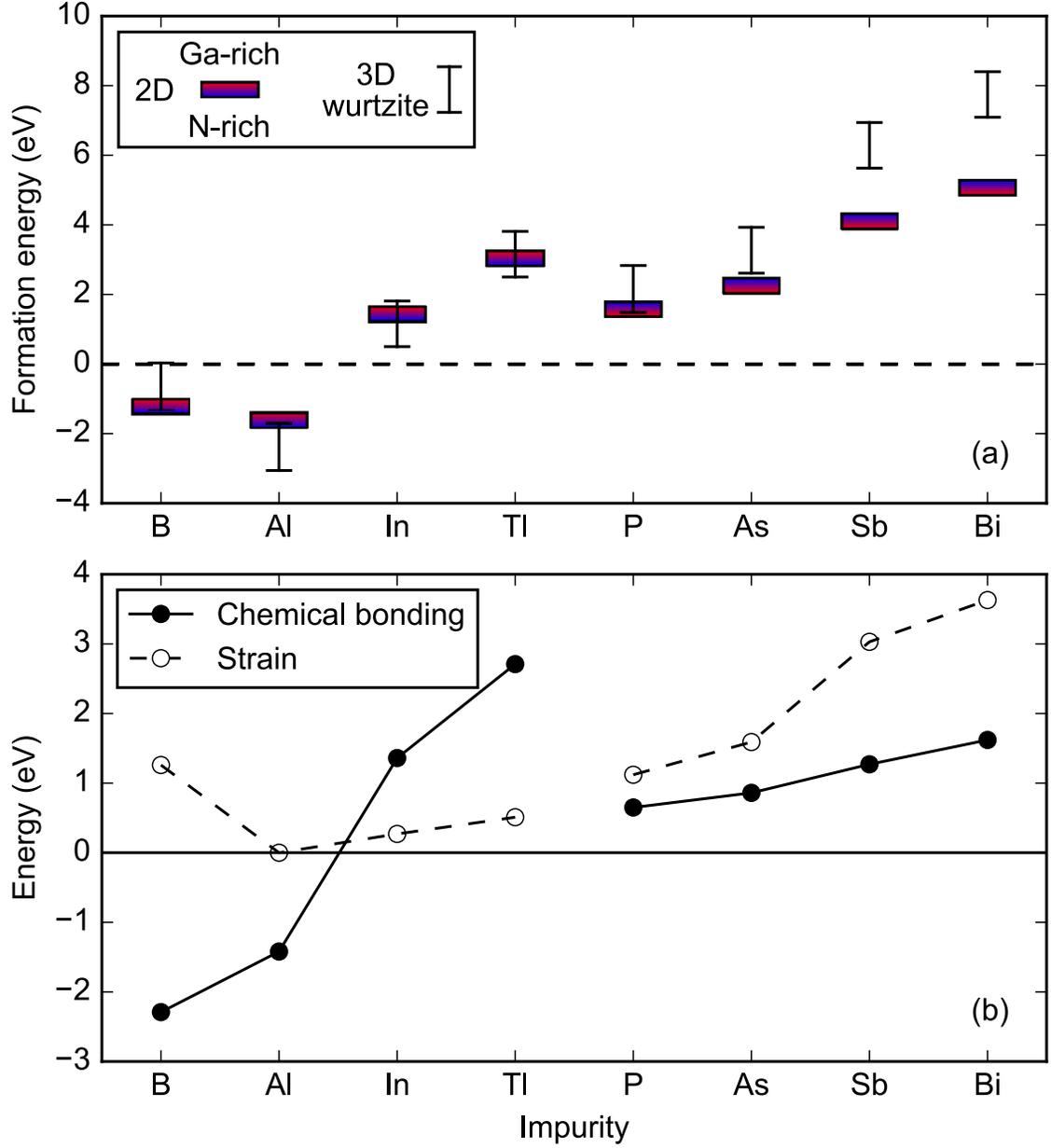}
	\caption{(a) Formation energies of substituted isoelectronic III-V elements in GaN . The range of values for a given element is associated with Ga or N rich growth conditions. (b) The formation energies are subdivided into the strain energy and change in the chemical energy for 2D-GaN. The chemical trend for group-III elements is governed by the chemical energy, while the strain energy dominates for group-V elements.}
	\label{FIG:formation_enthalpy}
\end{figure}

The feasibility of incorporating the alloying element into GaN is assessed by calculating defect formation energies due to individual impurity elements following the method outlined in Ref.~\onlinecite{Freysoldt_RMP_86_2014}. The defect formation energy (here demonstrated for a general group-III substitution $X_\text{Ga}$) is calculated using the DFT total energy of the supercell with an impurity as well as the respective host structure
\begin{eqnarray}\label{Eq:Ef}
	E_\text{f}[X_1\text{Ga$_{63}$N$_{64}$}] &=& E_\text{tot}[X_1\text{Ga$_{63}$N$_{64}$}] - E_\text{tot}[\text{Ga$_{64}$N$_{64}$}] - \nonumber \\
	& & \mu[X] + \mu[\text{Ga}].
\end{eqnarray}
The DFT total energies of bulk naturally occurring structures are taken to represent the chemical potential of impurities $\mu[X]\approx E_\text{tot}[X_\text{bulk}]$. The gallium chemical potential is confined within the range
\begin{equation}\label{Eq:mu[Ga]_2}
	E_\text{tot}[\text{Ga}_\text{bulk}]+\Delta H_\text{f}[\text{GaN}]  < \mu[\text{Ga}] < E_\text{tot}[\text{Ga}_\text{bulk}],
\end{equation}
where $\Delta H_\text{f}$ is the calculated formation enthalpy of GaN ($-0.42$ and $-1.29$~eV for the 2D hexagonal planar and bulk wurtzite structures, respectively) determined as
\begin{equation}\label{Eq:dH[GaAs]}
	\Delta H_\text{f}[\text{GaN}] \approx E_\text{tot}[\text{GaN}] - \frac{1}{2}E_\text{tot}[\text{N}_2] - E_\text{tot}[\text{Ga}_\text{bulk}].
\end{equation}
The upper/lower limit of $\mu[\text{Ga}]$ corresponds Ga rich/poor growth conditions.

The results of the defect formation energy calculations are presented in Fig.~\ref{FIG:formation_enthalpy}a. Among all III-V elements only B or Al substitutions are energetically favourable. Indium has a limited solubility in the bulk GaN \cite{Ho_APL_69_1996}, and the same is expected for 2D-GaN due to the similar $E_\text{f}$ values. The energy penalty associated with the incorporation of a group-V element in 2D-GaN is systematically lower than in the bulk. Due to the relaxed strain conditions in monolayers,  2D materials may generically be more easily alloyed than their bulk counterparts. Phosphorus has a similar defect formation energy to indium in 2D-GaN, which suggest the possibility for practical realization of dilute metastable 2D-GaN$_{1-x}$P$_x$. 

To shed light on the physics behind observed chemical trends in the formation energies of the group III-V substitutions, we decomposed the $E_\text{f}$ values into the strain energy and chemical energy components. The strain and chemical energies are defined as
\begin{subequations}
 \label{EQN:strain_energy-and-bonding_energy}
 \begin{eqnarray}
  E_{\text{strain}} &=& E_{\text{tot}} [X_1\text{Ga}_{63}\text{N}_{64}] - 63 E_{\text{tot}} [\text{GaN}] -  E_{\text{tot}} [X\text{N}],\label{EQN:strain_energy}
 \\
  \Delta E_{\text{chem}} &=& \Delta H_\text{f} [X\text{N}] - \Delta H_\text{f} [\text{GaN}].\label{EQN:bonding_energy}
 \end{eqnarray}
\end{subequations}
It can be shown that $E_{\text{strain}}+\Delta E_{\text{chem}}$ correspond to the upper limit of $E_\text{f}$ in Eq.~(\ref{Eq:Ef}). The values of $E_{\text{strain}}$ and $\Delta E_{\text{chem}}$ attained are presented in Fig.~\ref{FIG:formation_enthalpy}b. In the case of boron the high strain energy is compensated by the strong chemical bonding. The incorporation of In and Tl is primarily limited by the weak chemical bonding. In contrast, the solubility of group-V elements is impeded by a high strain energy due to the large variance in the atomic size from the host N.

\begin{figure}
	\includegraphics[width=1\columnwidth]{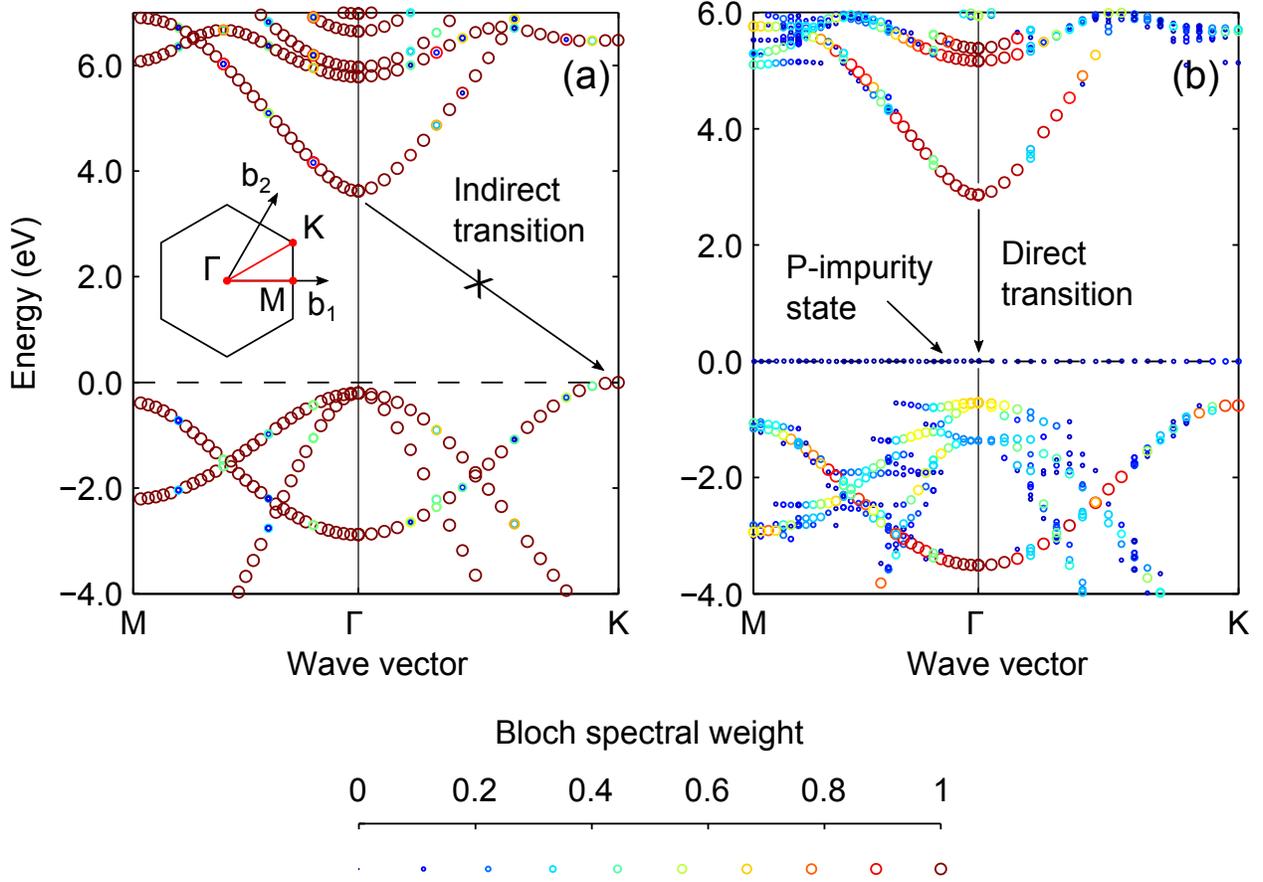}
	\caption{Effective band structure of a hexagonal monolayer of (a) Ga$_{25}$N$_{25}$ and (b) Ga$_{25}$N$_{24}$P$_1$ supercells obtained with HSE06 functional. Phosphorus introduces a localized state (without a well-defined Bloch character) above the valence band of the host GaN. A partial $\Gamma$-character of this impurity-like state creates a possibility for optical transitions in otherwise indirect band gap 2D-GaN. The origin of the energy scale is set at the Fermi energy.}
	\label{FIG:bandstructures}
\end{figure}

The analysis of the defect formation energies and the band gap reduction efficiency points to a dilute 2D-GaN$_{1-x}$P$_x$ alloy as a favourable candidate for 2D-GaN-based light emitters in the visible spectrum. With this purpose in mind we investigated optoelectronic properties of GaN$_{0.96}$P$_{0.04}$ using the Heyd-Scuseria-Ernzerhof  (HSE06) screened hybrid functional \cite{Heyd_JCP_118_2003,Krukau_JCP_125_2006} intended to overcome the band gap underestimation inherent to DFT semilocal exchange correlation functionals. We have estimated the lattice constant of the GaN$_{1-x}$P$_x$ alloy using Vegard's law. Reduction of the band gap by 0.57~eV is observed with respect to the host 2D-GaN. The effective band structure unfolded to a primitive Brillouin zone using \texttt{fold2Bloch} package \cite{Rubel_PRB_90_2014} is shown in Fig.~\ref{FIG:bandstructures}. Phosphorous introduces a dispersionless state approximately 0.6~eV above the host valence band. The lack of a well-defined Bloch character is consistent with the strong localization at the VBE for all group-V impurities (Table~\ref{TABLE:born_eff_loc}).

A non-zero dipole transition matrix element is expected due to a direct optical transition between P-state and the conduction band edge, whose Bloch character at $\Gamma$-point is well preserved. A similar strategy is employed in green LEDs based on the bulk dilute ($\sim2\times10^{18}$~cm$^{-3}$) nitrogen-dopped GaP \cite{Dapkus_JAP_45_1974}. Although 2D-GaN$_{0.96}$P$_{0.04}$ is an optically active material, the  dipole matrix element for a direct transition in Fig.~\ref{FIG:bandstructures}b is two orders of magnitude weaker than the $\Gamma_\text{c}\rightarrow\Gamma_\text{v}$ matrix element in the planar 2D-GaN. This result can be attributed to a weak $\Gamma$-character of the P-related state as evident from the unfolded band structure. Thus, the intrinsically direct band gap of a few-layered 2D-GaN \cite{Balushi_NM_15_2016} makes it an attractive alternative to the planar structures for optical emitters. One would expect the behaviour of impurities in a few-layered 2D-GaN to fall between our results for the planar monolayer 2D-GaN and the bulk GaN depending on the layer thickness.

%
%

\section{Conclusion}

Isoelectronic substitutions into a host monolayer 2D-GaN were explored to achieve  optical emission with a tuneable wavelength in the visible spectrum. Low-dimensional materials may generally be less strenuous to alloy than bulk materials due to fewer geometrical constraints. Incorporation of phosphorous among other III-V elements enables a much more efficient band gap reduction than indium while featuring similar impurity formation energies. The band gap reduction of the order \mbox{$-100\ldots-\!300$~meV} per \%P in dilute 2D-GaN$_{1-x}$P$_x$ is mediated by formation of a localized state above the valence band edge of the host 2D-GaN. Concurrently, the conduction band remains resistant to perturbations irrespective of the substitutional element. The phosphorous impurity state facilitates a direct optical transition in otherwise indirect planar monolayer 2D-GaN, which becomes possible due to the electronic localization and the associated relaxation of the wave vector selection rule.

%
%
\begin{acknowledgments}
Authors would like to thank Maciej Polak (Wroclaw University of Science and Technology) and Alex Pofelski (McMaster University) for critical reading and valuable discussions. Funding was provided by the Natural Sciences and Engineering Research Council of Canada under the Discovery Grant Program RGPIN-2015-04518. C.P. would like to acknowledge support by the Ontario Graduate Scholarship. This work was performed using computational resources of the Thunder Bay Regional Research Institute, Lakehead University, and Compute Canada (Calcul Quebec).
\end{acknowledgments}

%
%

%

\end{document}